\documentclass[twocolumn, twocolappendix]{aastex631}

\received{09 April 2026}
\revised{29 May 2026}
\accepted{5 June 2026}

\usepackage{graphicx}
\usepackage{amsmath}
\usepackage{xspace}
\usepackage{tabularx}
\usepackage[hang,flushmargin]{footmisc}
\usepackage{multirow}

\usepackage{xcolor}

\begin{document}

\title{Flux-cube reconstruction from slitless spectroscopy}

\correspondingauthor{M.\ Griggio}
\email{mgriggio@stsci.edu}

\author[0000-0002-5060-1379]{M.\ Griggio}
\affiliation{Space Telescope Science Institute,
3700 San Martin Drive,
Baltimore, MD 21218, USA}

\author[0000-0003-0894-1588]{R.\ E.\ Ryan\ Jr.}
\affiliation{Space Telescope Science Institute,
3700 San Martin Drive,
Baltimore, MD 21218, USA}

\author[0000-0003-3382-5941]{N.\ Pirzkal}
\affiliation{Space Telescope Science Institute,
3700 San Martin Drive,
Baltimore, MD 21218, USA}

\author[0000-0002-9888-2704]{T.\ L.\ Astraatmadja}
\affiliation{Space Telescope Science Institute,
3700 San Martin Drive,
Baltimore, MD 21218, USA}

\author{S.\ Casertano}
\affiliation{Space Telescope Science Institute,
3700 San Martin Drive,
Baltimore, MD 21218, USA}

\author[0000-0002-4436-4661]{S.\ Perlmutter}
\affiliation{E.O. Lawrence Berkeley National Laboratory,
1 Cyclotron Rd.,
Berkeley, CA, 94720, USA}
\affiliation{Department of Physics, University of California Berkeley, 366 LeConte Hall MC 7300, Berkeley, CA 94720, USA}

\author[0000-0001-5402-4647]{D.\ Rubin}
\affiliation{Department of Physics and Astronomy,
University of Hawai`i at M{\=a}noa,
Honolulu, HI, 96822}
\affiliation{E.O. Lawrence Berkeley National Laboratory,
1 Cyclotron Rd.,
Berkeley, CA, 94720, USA}

\author{G.\ Aldering}
\affiliation{E.O. Lawrence Berkeley National Laboratory, 1 Cyclotron Rd., Berkeley, CA, 94720, USA}

\author[0000-0002-7566-6080]{J.\ M.\ DerKacy}
\affiliation{Space Telescope Science Institute, 3700 San Martin Drive, Baltimore, MD 21218, USA}

\author[0000-0003-2238-1572]{O.\ D.\ Fox}
\affiliation{Space Telescope Science Institute, 3700 San Martin Drive, Baltimore, MD 21218, USA}

\author[0000-0002-6652-9279]{A.\ S.\ Fruchter}
\affiliation{Space Telescope Science Institute, 3700 San Martin Drive, Baltimore, MD 21218, USA}

\author[0000-0002-1296-6887]{L.\ Galbany}
\affiliation{Institute of Space Sciences (ICE-CSIC), Campus UAB, Carrer de Can Magrans, E-08193 Barcelona, Spain}
\affiliation{Institut d'Estudis Espacials de Catalunya (IEEC), E-08860 Castelldefels, Barcelona, Spain}

\author[0000-0002-0476-4206]{R.\ Hounsell}
\affiliation{University of Maryland Baltimore County, 1000 Hilltop Cir, Baltimore, MD 21250, USA}
\affiliation{NASA -- Goddard Space Flight Center, 8800 Greenbelt Rd, Greenbelt, MD 20771, USA}

\author[0009-0006-9538-4781]{A.\ M.\ Isaacs}
\affiliation{School of Physics and Astronomy, University of Minnesota, 116 Church Street SE, Minneapolis, MN 55455, USA}

\author[0000-0003-3142-997X]{P.\ L.\ Kelly}
\affiliation{School of Physics and Astronomy, University of Minnesota, 116 Church Street SE, Minneapolis, MN 55455, USA}

\author[0000-0003-3221-0419]{R.\ S.\ Kessler}
\affiliation{Department of Physics, University of Chicago, 5720 So. Ellis Ave., KPTC 201, Chicago, IL 60637-1434, USA}
\affiliation{Fermi National Accelerator Laboratory, P.O. Box 500, Batavia, IL 60510-0500, United States}

\author[0000-0002-1873-8973]{B.\ M.\ Rose}
\affiliation{Department of Physics and Astronomy, Baylor University, One Bear Place \#97316, Waco, TX 76798-7316, USA}

\author[0009-0005-9470-0765]{J.\ Roychowdhury}
\affiliation{Department of Physics, Duke University, Durham, NC 27708, USA}

\author[0000-0003-2764-7093]{M.\ Sako}
\affiliation{Department of Physics and Astronomy, University of Pennsylvania, 209 South 33rd Street, Philadelphia, PA 19104, USA}

\author[0000-0002-4934-5849]{D.\ M.\ Scolnic}
\affiliation{Department of Physics, Duke University, Durham, NC 27708, USA}

\author[0000-0002-7756-4440]{L.-G.\ Strolger}
\affiliation{Space Telescope Science Institute, 3700 San Martin Drive, Baltimore, MD 21218, USA}

\collaboration{100}{and the Roman Supernova Cosmology Project Infrastructure Team\footnote{\url{https://www.romansnpit.com}}}

\begin{abstract}
Slitless spectroscopy enables efficient, large-area surveys without target pre-selection, yet it faces challenges from source blending, higher noise, and lost spatial-spectral information. We present an advanced, non-parametric, data-driven algorithm that leverages multiple dispersion angles to reconstruct three-dimensional flux distributions, providing low-resolution Integral Field Unit (IFU) capabilities from slitless data. By treating each pixel as an independent element, our method naturally handles source confusion without requiring prior assumptions regarding redshifts, templates, or model libraries. We validate the algorithm using simulated Roman Space Telescope wide-field slitless spectroscopy images that are equivalent to what is expected from the High-Latitude Time-Domain Survey. First, we demonstrate that a host-galaxy model reconstructed from multiple dispersion angles can be used to accurately subtract host light from a transient, recovering a Type Ia supernova spectrum with minimal bias. Second, we showcase a high-fidelity flux-cube reconstruction of a complex galaxy, successfully measuring the redshift and recovering continuum, emission, and absorption features. This approach highlights the potential of multi-dispersion-angle slitless data to provide spatially resolved spectral information in a non-parametric way, which is traditionally accessible only with integral field spectroscopy, opening a new window into large, unbiased, and spatially-resolved studies of galaxy evolution.
\end{abstract}


\section{Introduction} \label{sec:intro}

Slitless spectroscopy is an observational technique in which a dispersive element, such as a grism or prism, is inserted into the optical path without the use of a physical aperture (e.g., slit or fiber), allowing spectra to be obtained simultaneously for all sources in the field of view. This technique is well-suited for wide-field imagers, where the absence of a physical aperture provides unbiased and efficient spectral coverage for large areas. However, the absence of a slit poses significant challenges in data analysis, primarily due to adjacent spectra overlapping as well as significantly higher sky background compared to slit-based spectroscopy. These effects require precise calibration and advanced extraction techniques to resolve overlapping signals and accurately extract one-dimensional spectra \citep{2001STECF..29....5P,axe09,ryan18,grizli}.  

Despite these limitations, slitless spectroscopy has proven to be a transformative tool for probing the high-redshift Universe. Early on, the Advanced Camera For Surveys (ACS) G800L grism on the Hubble Space Telescope (HST) was foundational for time-domain cosmology, providing a sample of spectroscopically confirmed, host-independent Type Ia supernovae to constrain the cosmic expansion history \citep{2003ApJ...589..693B, 2004ApJ...600L.163R}. In the Hubble Ultra Deep Field, the Grism ACS Program for Extragalactic Science (GRAPES) survey \citep{2004ApJS..154..501P} utilized this mode to conduct an unbiased, continuum-selected census of high-redshift Ly$\alpha$ emitters \citep{2005ApJ...626..666M, 2005ApJ...621..582R} and identify low-luminosity AGN populations \citep{2007AJ....134..169X}. Building on the pioneering wide-field slitless spectroscopy (WFSS) efforts with Near Infrared Camera and Multi-Object Spectrometer (NICMOS) \citep{1999ApJ...520..548M}, GRAPES was the first survey to strategically employ multiple dispersion angles to disentangle overlapping spectra and vet detected emission lines. While the NICMOS pilot was restricted to a 51$^{\prime\prime}$$\times$51$^{\prime\prime}$ field of view, GRAPES leveraged the significantly larger 202$^{\prime\prime}$$\times$202$^{\prime\prime}$ footprint of ACS, establishing a multi-orientational strategy that became critical for subsequent WFSS surveys. This approach enabled the study of galaxy formation and evolution at intermediate redshifts \citep{2005ApJ...626..680D, 2006ApJ...636..115P} and probed Galactic structure through the detection of faint M and L dwarfs \citep{2005ApJ...622..319P}.

The importance of this multiple-dispersion-angle strategy is evidenced in the design of the instruments on Euclid/NISP and NIRCam and NIRISS on the \textit{James Webb Space Telescope} (JWST), where multiple grisms allow for instantaneous changes in dispersion direction without requiring telescope re-orientation.  By leveraging enhanced sensitivity and wavelength coverage, the JWST has pushed the limits of slitless spectroscopy, enabling the identification of high-redshift active galactic nuclei, emission line measurements across a wide range of redshifts, and the study of feedback and the mass-metallicity relation in faint dwarf galaxies \citep{2023ApJ...946L..13F,2023ApJ...946L..14K,2023MNRAS.525.2864O,2024ApJ...969...90P,2025arXiv251115792M}.

The complexity of slitless data has driven the development of various extraction methodologies over the past two decades, evolving from early geometric tools like \texttt{aXe} \citep{2001STECF..29....5P} to more sophisticated two-dimensional modeling techniques. Modern techniques have further refined the management of spectral overlap by shifting toward more sophisticated scene-based analysis. Beyond the \texttt{aXe} methodology, \texttt{Grizli} \citep{grizli} utilizes spectral models (whether empirical or synthetic object templates or polynomial models) to estimate and subtract contamination from neighboring sources, potentially offering an improvement over models based purely on photometry. In contrast, \texttt{LINEAR} \citep{ryan18} establishes a comprehensive forward model of the scene to create a system of sparse linear equations, which can then be solved using specialized algorithms to extract deblended spectra.
Complementary to these are the \texttt{MAP2D} and \texttt{EM2D} algorithms \citep{2017wfc..rept....1P,2018ApJ...868...61P,2024ApJ...969...90P}, which improved upon one-dimensional extraction frameworks to identify emission line regions of galaxies in the WFSS data using multiple dispersion angles and to make two-dimensional maps of these regions.

In principle, these forward-modeling frameworks can mitigate basic self-confusion within extended sources by segmenting a single galaxy into a discrete, macroscopic spatial sub-components (e.g., separating a bulge from a disk). However, implementations that utilize this segmentation approach rely on template-based fitting for each sub-component \citep[e.g.,][]{2024MNRAS.532..577E}, restricting the extracted spectra to predefined empirical or synthetic models. Conversely, while the mathematical foundation for a completely non-parametric, pixel-level linear reconstruction was pioneered by \cite{ryan18}, it was never practically demonstrated or validated for reconstructing extended sources. Crucially, previous linear-inversion implementations omitted the point spread function from the forward-modeling, severely limiting the ability to reconstruct the full spatial-spectral morphology of an object. Recovering this full, continuous spatial-spectral information is essential when a localized feature of interest, such as a transient or a localized star-forming region, is embedded within a host galaxy, a problem that is essentially akin to deblending independent sources within a single, extended object \citep{neveu}. 

While multiple dispersion angles effectively mitigate external contamination from neighboring stars or galaxies \citep{ryan18}, standard linear-reconstruction algorithms often remain hampered by an inability to fully isolate the underlying host component from the feature of interest \citep{joshi22}. This is particularly difficult in extended galaxies with complex spatial-spectral morphologies, where disparate regions (e.g., a quiescent nucleus versus clumpy, star-forming spiral arms) possess distinct spectral energy distributions (SEDs). Because different dispersion angles produce unique projections of this three-dimensional flux-density cube onto the two-dimensional detector, the result is ``self-contamination''---where the spectral traces of distinct internal subregions overlap in a way that is difficult to disentangle using traditional extraction methods.  To address this, \citet{tri} demonstrated that data taken at multiple orientations can be used to marginalize over spatial-spectral information to effectively ``null out'' the two-dimensional background spectra of extended host objects. However, the efficacy this technique remains strictly limited when the test orientation deviates significantly from the specific orientations used to train the spatial model.

In this paper, we extend the forward-modeling framework to extract three-dimensional flux cubes from extended sources by leveraging the diverse spatial information provided by multiple dispersion angles. We use the Roman High-Latitude Time-Domain Survey as a primary case study, where the high number and diversity of available dispersion angles ($\gtrsim$\,72) offers a unique opportunity to reconstruct the intrinsic spatial-spectral structure of galaxies. By applying our technique to simulated Roman images, we demonstrate its capability to effectively disentangle overlapping spatial and spectral features, transforming traditional slitless spectroscopy into a tool for spatially resolved analysis.

The paper is organized as follows. Section\,\ref{sec:wfss} and \ref{sec:lin} present the formalism used to build the forward model of a scene. Section\,\ref{sec:rec} describes how we invert the problem and find an optimal solution. In Section\,\ref{sec:app} we illustrate two scientific applications of our method using Roman simulated images. We detail our implementation of the algorithm and provide some benchmarks in Section\,\ref{sec:alg}, and we conclude the paper in Section\,\ref{sec:concl}.

\section{The Image Formation}
\label{sec:wfss}

\begin{figure*}
    \centering
    \includegraphics[trim=0 80 0 0,clip,width=.75\textwidth]{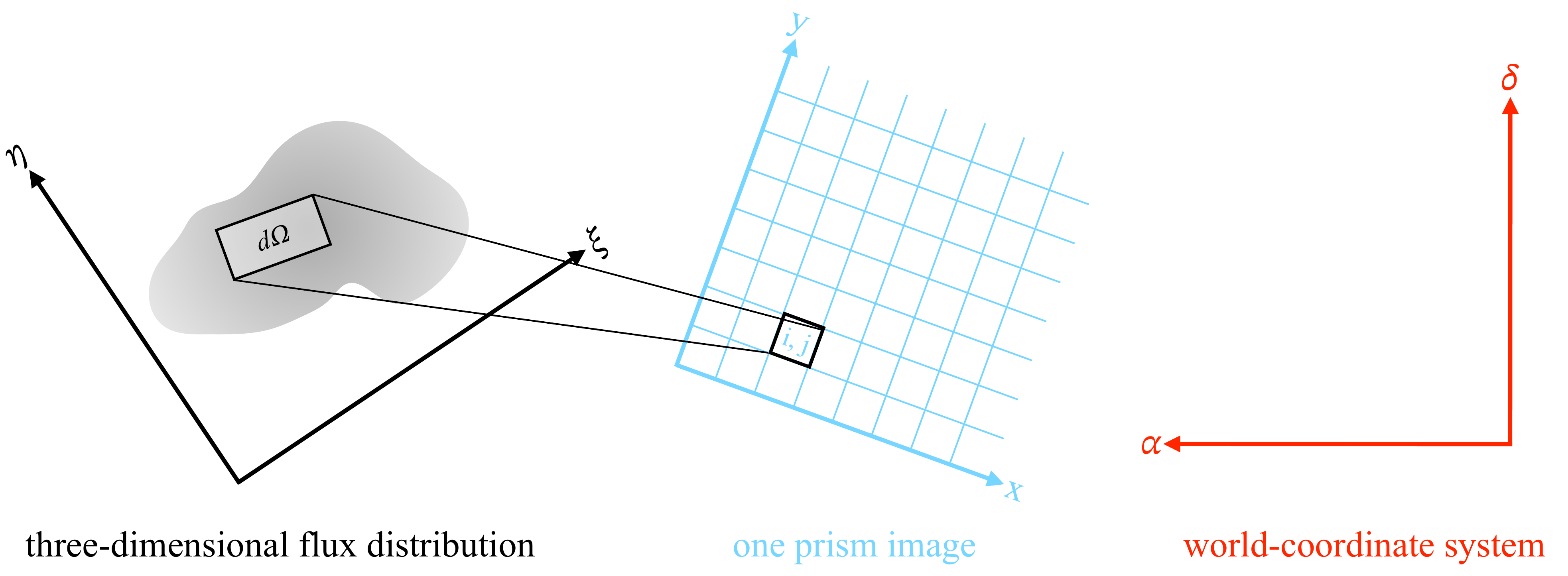}
    \caption{
    Left: illustration of the astronomical scene and its coordinate system $(\xi, \eta)$. The grey shape represents an extended source.
    Middle: the detector coordinate system $(x,y)$ and its pixelated representation $(i,j)$. The black lines represent the area subtended by the pixel $(i,j)$ in the astronomical scene, $d\Omega$.
    Right: equatorial reference system. In principle, $(\alpha,\delta)$ are not aligned with $(\xi,\eta)$ nor $(x,y)$. However, since $(\xi,\eta)$ are arbitrary, we typically choose them such as they are aligned to $(\alpha,\delta)$. The transformations between $(x,y)$ and $(\alpha,\delta)$ are given by the WCS.}
    \label{fig:schematic}
\end{figure*}

An astronomical scene can in general be represented by a three-dimensional flux distribution \citep[a spatial map of flux densities at different wavelengths; see, e.g.,][]{2016SPIE.9910E..16P,neveu,tri}, which can be rendered on a detector as a {\it direct image} (as with standard imaging filters) or a {\it spectral image} (as with spectroscopic gratings). If we denote the flux distribution as $\mathcal{C}(\xi,\eta,\lambda)$, the flux $\mathrm{F}$ in a pixel $(i,j)$ of a direct image is given by
\begin{equation}
    \label{eq:1}
    \mathrm{F}_{ij} = \int\!\!\!\int \mathcal{P}_{ij}(\lambda)\ast\left[(\mathcal{T}_{ij}(\lambda)\,\mathcal{C}(\xi,\eta,\lambda)\right] \,\lambda {\rm d}\lambda {\rm d}\Omega,
\end{equation}
where $d\Omega$ is the solid angle spanned by the area of the pixel $(i,j)$ projected into the $(\xi, \eta)$ coordinate space (see Fig.\,\ref{fig:schematic}) and $\mathcal{P}_{i,j}(\lambda)$ and $\mathcal{T}_{i,j}(\lambda)$ are the spatially varying, wavelength-dependent PSF and transmission curve, respectively.

The flux in a given pixel of a spectral image is obtained in a similar way, by adjusting the spatial coordinates by the respective functions of wavelength, i.e. substituting in Eq.\,\ref{eq:1}:
\begin{eqnarray*}
    &\xi& \rightarrow \xi + \Delta\xi_{i,j}(\lambda)\\
    &\eta&  \rightarrow \eta + \Delta\eta_{i,j}(\lambda),
\end{eqnarray*}
where $\Delta\xi$ and $\Delta\eta$ represent the \textit{optical model} that describes the dispersion, which depend in general on the position on the detector $(i,j)$ and is intrinsic to the instrument. The optical model must be calibrated by on-sky observations, which we assume will be provided.

To determine the limits of the spatial integrals we need to know the transformations between the scene $(\xi,\eta)$ coordinates and the detector $(i,j)$ pixel space, which depend on the known properties of the instrument (e.g., pixel scale and distortion model) and the pointing of the telescope. 
The scene reference system $(\xi, \eta)$ is typically assumed to be a distortion-free, celestial system in gnomonic projection. Its axes can be set to arbitrary directions, but typically they are aligned to the axes of the equatorial coordinate system $(\alpha, \delta)$.
The transformations from the detector coordinates $(i,j)$ and $(\alpha, \delta)$ are facilitated by a system of coordinate transform equations encoded in the World Coordinate System (WCS) convention \citep{gre02, cal02}.

\section{Building the linear operator}
\label{sec:lin}

As described in the previous section, the three-dimensional flux distribution $\mathcal{C}$ is a continuous function in both the spatial and wavelength dimensions. However, since our instruments have a finite resolution, we can only reconstruct a ``sampled'' (or \textit{pixelated}) representation. We denote this sampled distribution with $\mathrm{C}$. In the following, we will use italic font for continuous functions and roman font for their pixelated representations.

In our framework, we represent an extended source by discretizing the continuous flux distribution $\mathcal{C}(\xi, \eta, \lambda)$ into a finite grid of spatial and spectral bins. Spatially, we approximate the scene as a collection of point sources, each located at the center of its respective spatial bin, with an SED equal to the total flux falling within that bin. In the limit of infinitely small bins, this representation converges to the continuous source distribution. This spatial discretization is coupled with a spectral sampling of the SEDs into discrete flux elements. The resulting sampled distribution, $\mathrm{C}(p, q, k)$, can be viewed as a ``flux-cube''---a series of two-dimensional images at sampled wavelengths $\lambda_k$. 

If we consider this discretized distribution, the integrals in Eq.\,\ref{eq:1} can be written as discrete summations, and the measured flux $\mathrm{F}_{i,j}$ can be cast as:
\begin{equation}
    \label{eq:2}
    \mathrm{F}_{i,j} = \sum_{p,q,k} \mathrm{A}_{i,j,p,q,k} \mathrm{C}_{p,q,k}.
\end{equation}
By mapping the detector pixel indices $(i,j)$ to a single observation index $m$ and the flux-cube indices $(p,q,k)$ to a single spatial-spectral element index $n$,
\begin{eqnarray*} \label{eqn:remap}
   (i,j)&\rightarrow& m \\
   (p, q, k) &\rightarrow& n    
\end{eqnarray*}
Eq.\,\ref{eq:2} is transformed into a system of linear equations:
\begin{equation}
    \label{eq:3}
    \mathrm{F}_m = \sum_{n} \mathrm{A}_{m,n} \mathrm{f}_n,
\end{equation}
where $\mathrm{F}_m$ and $\mathrm{f}_n$ are the elements of the flattened observation and source spectral vectors, respectively. In this form, each row $m$ of the linear operator $\mathbf{A}$ corresponds to a detector pixel and each column $n$ corresponds to a specific spatial-spectral element.\\

To determine the coefficients $\mathrm{A}_{i,j,p,q,k}$, we model the transformations of a discrete flux element as it propagates through the optical system. Each element in the flux-cube is treated as a point source; given the pointing of the telescope and the dispersion curve of the instrument, an element at index $(p,q,k)$ maps to a continuous detector coordinate $(x, y)_{p,q,k}$. In the pixel-based coordinate system, this position is expressed as $(i_0, j_0) + (\delta i, \delta j)_{p,q,k}$, where $(i_0, j_0)$ are the integer indices of the pixel and $(\delta i, \delta j)_{p,q,k} \in [-0.5, 0.5]$ represents the fractional sub-pixel displacement from the pixel center. The dependence of this displacement on the spectral index $k$ is governed by the dispersion law, while the dependence on $(p, q)$ is determined by the geometric mapping from scene to detector.\\

If no other effects were present, the total flux from a source element $\mathrm{C}_{p,q,k}$ would be captured entirely by the pixel $(i_0,j_0)$. However, the optical system introduces a point spread function (PSF) that spatially redistributes the flux. Because each flux element is treated as a point source, the contribution to the detector is modeled by centering a monochromatic PSF, evaluated at wavelength $\lambda_k$, at the projected coordinates $(x, y)_{p,q,k}$. While a PSF theoretically extends to infinity, numerical implementation requires truncation at a finite radius. For every pixel $(i,j)$ within this truncation boundary, the fraction of flux is calculated by integrating the PSF over the area of the pixel. These fractional values are then scaled by additional factors that depend on the pixel coordinates $(i,j)$ and the wavelength index $k$, including the transmission efficiency, the flat-field response, and the local pixel area. These resulting values constitute the coefficients of the operator $\mathbf{A}$. This procedure is repeated for every element in the discretized flux-cube to construct the complete forward model.\\

While Equation\,\ref{eq:3} describes the transformation of a flux-cube into a single spectral image observed at a particular pointing, the formalism is designed to scale with both the volume of the data and the complexity of the scene. The framework can be extended to multiple exposures by vertically stacking the observation vectors $\mathbf{F}$ and the corresponding forward operators $\mathbf{A}$. This stacking ensures that every pixel from every available image is represented as a unique row in the global system of equations. Furthermore, since each spatial-spectral cell is treated as an independent flux element regardless of the astrophysical source to which it belongs, additional sources or increased spatial resolution are accommodated by appending columns to the operator $\mathbf{A}$. This modularity allows the inclusion of an arbitrary number of flux elements by expanding the source vector $\mathbf{f}$. 

The final result is a large, sparse system of linear equations that simultaneously models the contribution of every discretized element of the scene across all available observations. The stability and uniqueness of the resulting inversion depend on the balance between the number of observational constraints (which is the number of rows: $m$) and the number of model parameters (which is the number of columns: $n$). To ensure the system remains overdetermined, we require an observational strategy that satisfies $m\!>\!n$.

Since the number of columns ($n$) is defined by the product of spatial pixels and spectral elements, this framework allows for a flexible trade-off between spatial and spectral resolution. For a fixed volume of input data, one can to some extent ``trade'' degrees-of-freedom by undersampling the spatial grid to achieve higher spectral sampling, or vice versa. This modularity enables the reconstruction to be tuned to the specific scientific requirements of the target while maintaining a numerically well-posed linear system.

\section{The flux-cube reconstruction}
\label{sec:rec}
Having establish the procedure to determine the linear operator $\mathbf{A}$, the problem of reconstructing the flux-cube from a set of slitless spectroscopy observations is just a matter of inverting $\mathbf{A}$. 
However, while $\mathbf{A}$ is a sparse matrix, its inverse is typically dense, and directly computing the inverse is generally not possible. Direct inversion is computationally more expensive and numerically less stable than decomposition-based direct solvers or iterative methods. Standard direct methods often become computationally prohibitive due to the \textit{fill-in}\footnote{Fill-in occurs when zero entries in the original matrix become non-zero during the factorization process.} of sparse matrices during factorization, leading to excessive memory and time requirements.

To overcome these limitations, iterative methods have been developed as a powerful alternative. These methods generate a sequence of approximate solutions that converge to the true solution without explicitly modifying the matrix $\mathbf{A}$, thus preserving its sparse structure, while improving the computational efficiency. Among the most effective iterative solvers for sparse least-squares problems are the Krylov subspace methods, which include \texttt{LSQR} \citep{1982ACMTM...8...43P} and \texttt{LSMR} \citep{2011SJSC...33.2950F}.

An additional challenge is that the system can be ill-conditioned, meaning that small perturbations in the data can lead to disproportionately large changes in the solution. To address ill-conditioning, the objective function is modified to include Tikhonov regularization:  $\left|\left|\mathrm{A}\mathrm{f} - \mathrm{F} \right|\right|^2 + \ell^2\left|\left|\mathrm{f}\right|\right|^2$. This choice of regularization introduces a damping parameter ($\ell$) to stabilize the solution by penalizing large solution norms and effectively damps the high-frequency structure in the solution. This regularized problem is efficiently solved by applying a suitable iterative solver to an equivalent augmented linear system. This approach avoids the numerical instability and computational cost associated with setting up the normal equations of the original problem. Instead, the solver operates on the augmented system, which is formulated as a standard least-squares problem:
\begin{equation}
    \min_{\mathrm{f}}
    \left|\left| 
    \begin{pmatrix}
    \mathrm{A}\mathrm{f} \\
    \ell \mathrm{I}
    \end{pmatrix} 
    -
    \begin{pmatrix}
    \mathrm{F} \\
    0
    \end{pmatrix}
    \right|\right|^2.
\end{equation}
By working with this augmented system, we effectively solve the damped problem iteratively, preserving the sparsity of $\mathbf{A}$ and providing a robust solution even when the original matrix is ill-conditioned.  

The choice of the damping parameter $\ell$ is crucial for balancing the trade-off between solution accuracy and stability. A very small damping parameter leads to solutions that are highly sensitive to noise in the data, potentially resulting in large, oscillatory, or physically unrealistic parameter estimates. Conversely, an excessively large damping parameter will heavily penalize large parameter changes, effectively shrinking the solution towards zero or some prior, thereby biasing the results and potentially preventing the algorithm from converging to the true underlying parameters. There have been many heuristic schemes for choosing the damping parameter: the L-curve analysis \citep{hansen92}, generalized cross-validation \citep{golub}, or minimizing correlations in the residual images \citep{tri}.\looseness=-1

In this framework, we follow the convention described by \citet{ryan18} to normalize the damping parameter by the Frobenius norm of the system matrix $\mathbf{A}$, such that $\ell = \hat{\ell} \|\mathbf{A}\|_{\rm F}$. This approach ensures that the relative damping factor $\hat{\ell}$ is dimensionless and scale-invariant, providing a consistent metric regardless of the physical units or the size of the system. Ultimately, the ideal choice of $\hat{\ell}$ is problem-dependent and requires careful consideration of the specific application, noise characteristics, and desired properties of the solution.

\section{Applications}
\label{sec:app}

In this section, we showcase two applications of our method by simulating observations that mimic the expected dataset of the \textit{Roman} HLTDS. We specifically adopt the observational strategy of the HLTDS as our primary test case, as its multi-epoch, multi dispersion angle design provides an ideal environment to evaluate the robustness of our reconstruction algorithm. 

To evaluate our method within the context of the HLTDS, we utilize simulated images generated using \texttt{Ilia}\footnote{\url{https://gitlab.com/astraatmadja/Ilia}, developed by T.\ Astraatmadja.}. These simulations start from truth flux-cubes derived from the Vela-Sunrise dataset \citep[][Ryan et al., in prep.]{vela19}, allowing us to create morphologically complex simulated galaxies with spatially varying SEDs. All simulations include a sky background of $\sim$\,1.4\,$e^-/\rm{s}$.

\subsection{The Roman High-Latitude Time-Domain Survey} \label{sec:hltds}

The \textit{Nancy Grace Roman Space Telescope} is the next flagship mission by NASA, currently on schedule to launch in September 2026. The telescope has a 2.4\,m mirror and its primary instrument, the Wide-Field Instrument (WFI), is composed of 18 4k\,$\times$\,4k HgCdTe detectors \citep{2020JATIS...6d6001M} with a pixel scale of $\sim\!0.11^{\prime\prime}$~pix$^{-1}$ that subtends $\sim$\,0.28\,deg$^2$. This pixel scale is similar to that of the Wide-Field Camera 3 IR channel (WFC3/IR) onboard \textit{HST}; however, the WFI has an instantaneous field-of-view $\sim\!215$ times larger than WFC3/IR. The WFI supports direct imaging through a set of eight broadband filters that have a total wavelength coverage of approximately $0.5-2.3~\mu$m. Additionally, WFI has two slitless spectroscopic elements: the prism (P120; \citealt{2025JATIS..11b5001E}), which covers $\sim\!0.75-1.80\,\mu$m with a resolution of $R\!\sim\!100$ and a grism (G150), which covers $\sim\!1.0-1.93\,\mu$m with $R\!\sim\!600$.

The HLTDS is one of the core-community surveys and has the primary objective of precisely measuring the expansion history of the Universe by discovering and characterizing Type Ia supernovae (SNe Ia). The survey has a northern (ELAIS-N1) and southern (EDFS) field that are divided into wide and deep tiers. The wide tier will cover a total of $\sim$\,18\,deg$^2$, while the deep tier will cover $\sim$6.5\,deg$^2$, both with a $\sim$\,10-day cadence for imaging. To maximize their observability, both fields are strategically located in the continuous viewing zones of the telescope, which are near the Ecliptic poles.

The survey will additionally include a spectroscopic component for the southern field that will provide detailed classification and redshift measurements for a subset of the supernovae, which is essential for accurate distance calculations. The wide and deep spectroscopic tiers will cover $\sim$\,4.5\,deg$^2$ and $\sim$\,0.56\,deg$^2$, respectively, and will both have a $\sim$\,5-day cadence. The total exposure time per visit in the wide and deep tiers are of 900\,s and 3600\,s, respectively, with eight dithers per visit. Given the intrinsic roll of the telescope of about 1\,deg per day, each field will be observed with multiple discrete dispersion angles, with about 5\,deg separation between each visit, resulting in approximately 72 different dispersion angles per field after one year of the survey. This variety in orientations, combined with a eight-dither strategy per visit, makes the HLTDS an excellent framework for testing our reconstruction method.

Beyond the core survey plan, the HLTDS will have a pilot survey in the initial phase to gather essential calibration data and to obtain an early measurement of the rate of high-redshift SNe Ia, which will provide critical feedback to nominal observing plan, and an extended component with longer cadence to provide temporal coverage for long-duration events. Additional details on the HLTDS design can be found in \cite{rotac25}, while for the expected survey yields see  \citep[\citealt{rose25}; see also][]{rubin25}.

\subsection{Host subtraction} \label{sec:hostsub}

\begin{figure}
    \centering
    \includegraphics[width=\columnwidth]{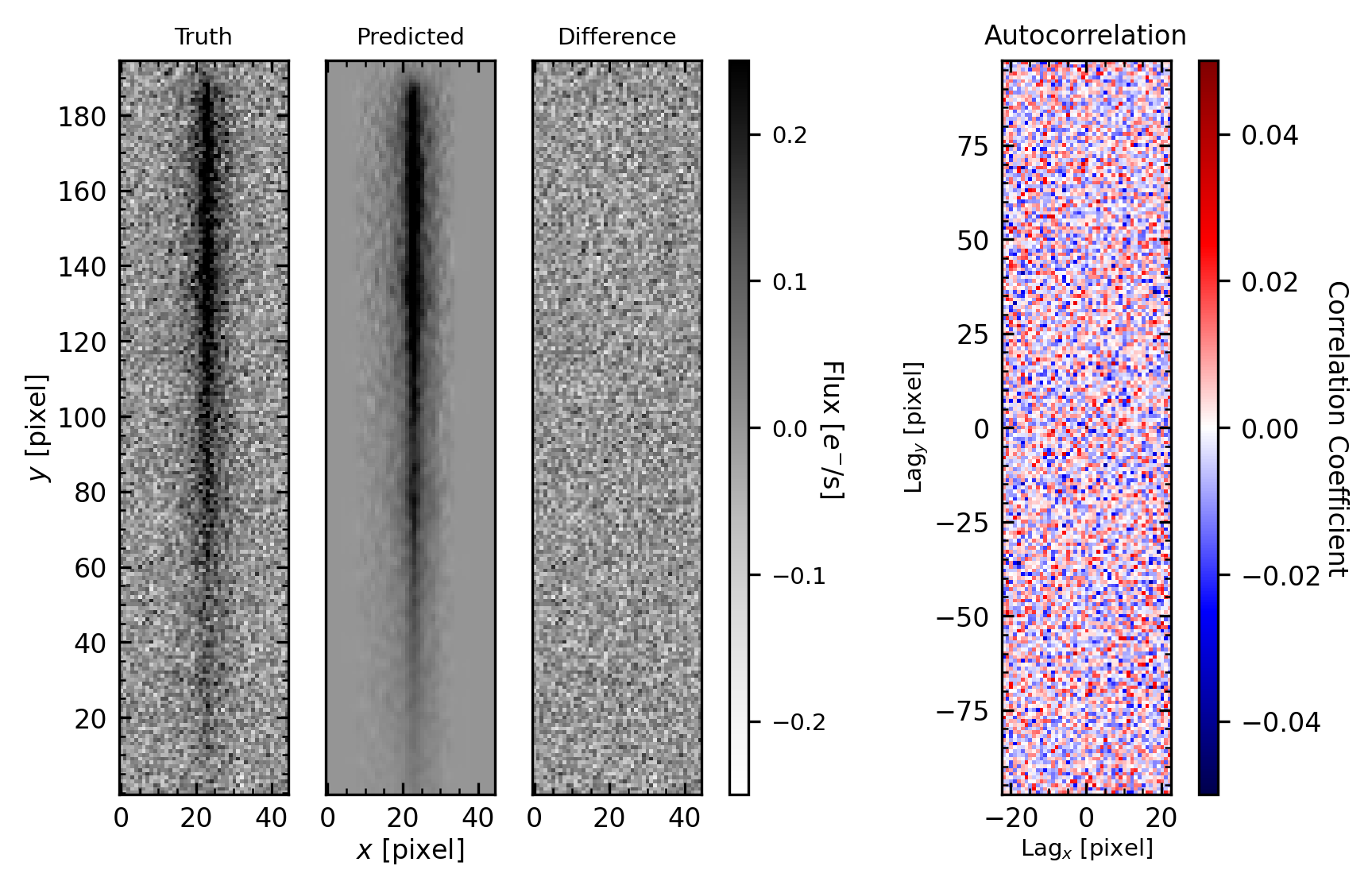}
    \caption{Left: simulated prism image of a $z=1$ galaxy at PA\,$=20^\circ$, image predicted using PA\,$=0^\circ,5^\circ,10^\circ,15^\circ$, and residuals. Right: two-dimensional autocorrelation of the residuals.}
    \label{fig:diff_acf}
\end{figure}

\begin{figure}
    \centering
    \includegraphics[width=\columnwidth]{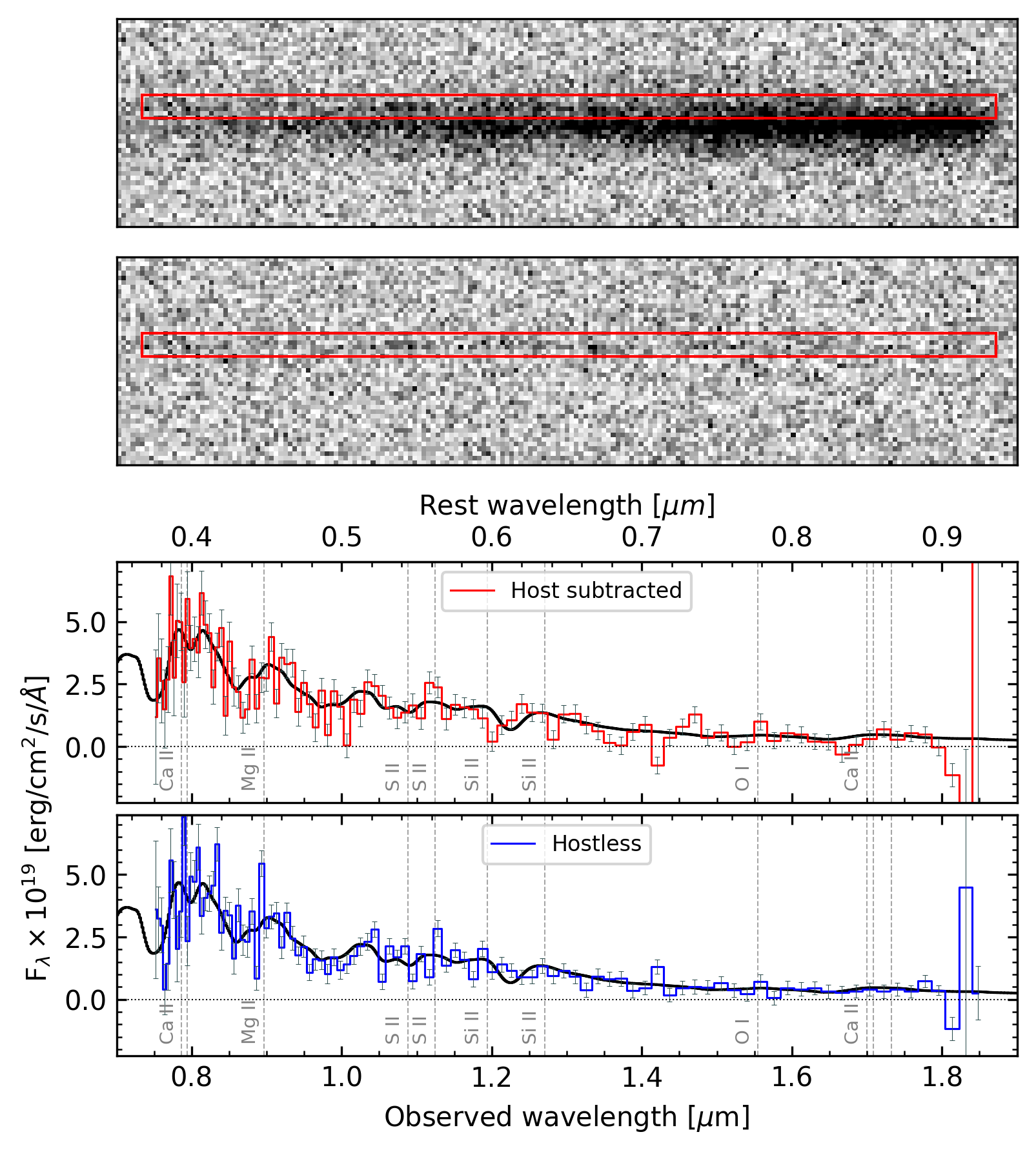}
    \caption{Top panels: spectral image of the same galaxy of Fig.\,\ref{fig:diff_acf} with a Type\,Ia SN injected in the region indicate by the red box, and the difference image obtained by subtracting the galaxy predicted trace of Fig.\,\ref{fig:diff_acf}. Bottom panels: extracted one-dimensional spectrum of the SN using our algorithm using the 8 dithered, host-subtracted images (red), and using the 8 dithered, hostless images of the SN. Key features which appear as P Cygni profiles in SNe~Ia are identified with vertical dashed lines.}
    \label{fig:extrIa}
\end{figure}

Slitless spectroscopy inherently presents a significant challenge for transient science: the spectral traces of the detected transients are unavoidably contaminated by the overlapping spectra of their respective host galaxies and neighboring sources. In this section, we present an example of decontamination using our linear reconstruction technique, focusing specifically on the challenge posed by the host galaxy contamination for a single transient event.

Our approach relies on using observations of the host galaxy where the transient is not present. For this demonstration, we consider a single transient appearing within its host galaxy, and we assume there are four distinct dispersion angles containing only the host. These four dispersion angles are separated by a $5^\circ$ step, with each dispersion angle comprising eight individual dithered exposures of 450\,s each, that is representative of the dataset provided after $\sim\!1$~month by the currently expected HTLDS observing strategy \citep{rotac25}. We consider the $z\!=\!1$ galaxy \texttt{Vela01} from Ryan et al. (in prep.).  These uncontaminated observations provide the necessary input for reconstructing the flux-cube of the host galaxy, combining the information across different dispersion angles to resolve the spatial and spectral structure of the extended source.

The decontamination process involves two steps: projection and subtraction. First, the host galaxy flux-cube model is projected into the dispersion angle of the image containing the transient (separated by a $5^\circ$ step from the previous). Second, this predicted two-dimensional spectral image of the host galaxy is subtracted from the observed image at that same dispersion angle. This subtraction isolates the spectral trace of the transient, significantly mitigating the host galaxy background contamination and enabling accurate spectroscopic extraction of the transient itself. As discussed in the previous sections, there are several techniques to determine the optimal damping parameter for the reconstruction. In this example, we determined the optimal value by uniformly spanning a range of damping parameters and selecting those that make the average of the squared two-dimensional autocorrelation of the residual image closest to zero \citep[similar to what done by][]{tri}, obtaining $\log_{10}(\hat{\ell})=-3$.

To better assess the residuals of the subtraction we simulated the fifth dispersion angle both with and without the transient. For the transient, we considered a $z\!=\!1$ Type Ia supernova spectrum, obtained form the template by \cite{hsiao}. In the left three panels of Fig.\,\ref{fig:diff_acf} we show the simulated image at the fifth dispersion angle with no transient (left), the predicted image (middle), and the difference image (right). The last panel on the right shows the two-dimensional autocorrelation of the residuals, which show no significant spatial residuals.

In Fig.\,\ref{fig:extrIa} we show the fifth dispersion angle image with the additional SN spectrum (top panel), with a red box indicating the SN trace. The second panel from the top shows the difference image obtained by subtracting the predicted two-dimensional spectral image of the galaxy (the same shown in Fig.\,\ref{fig:diff_acf}) from the image containing the transient. We then combined all the eight dithered, host-subtracted exposures to extract the one-dimensional spectrum of the SN. We performed the extraction with our code, in a similar way as in \cite{ryan18}, but with the addition of the wavelength dependent PSF. For comparison, we simulated 8 dithered exposures containing the SN light only, and performed the same extraction. The two one-dimensional spectra are plotted in Fig.\,\ref{fig:extrIa}. The black line represents the input spectrum, while the red and blue lines are the spectrum measured from the host-subtracted images and the one measured form the SN-only images, respectively. There is overall good agreement between the host-subtracted and hostless extractions; in particular we see no evidence that our subtraction resulted in biases in the SN spectrum.

\subsection{Spatially-resolved galaxy studies} \label{sec:resolved}

\begin{figure}
    \centering
    \includegraphics[width=\columnwidth]{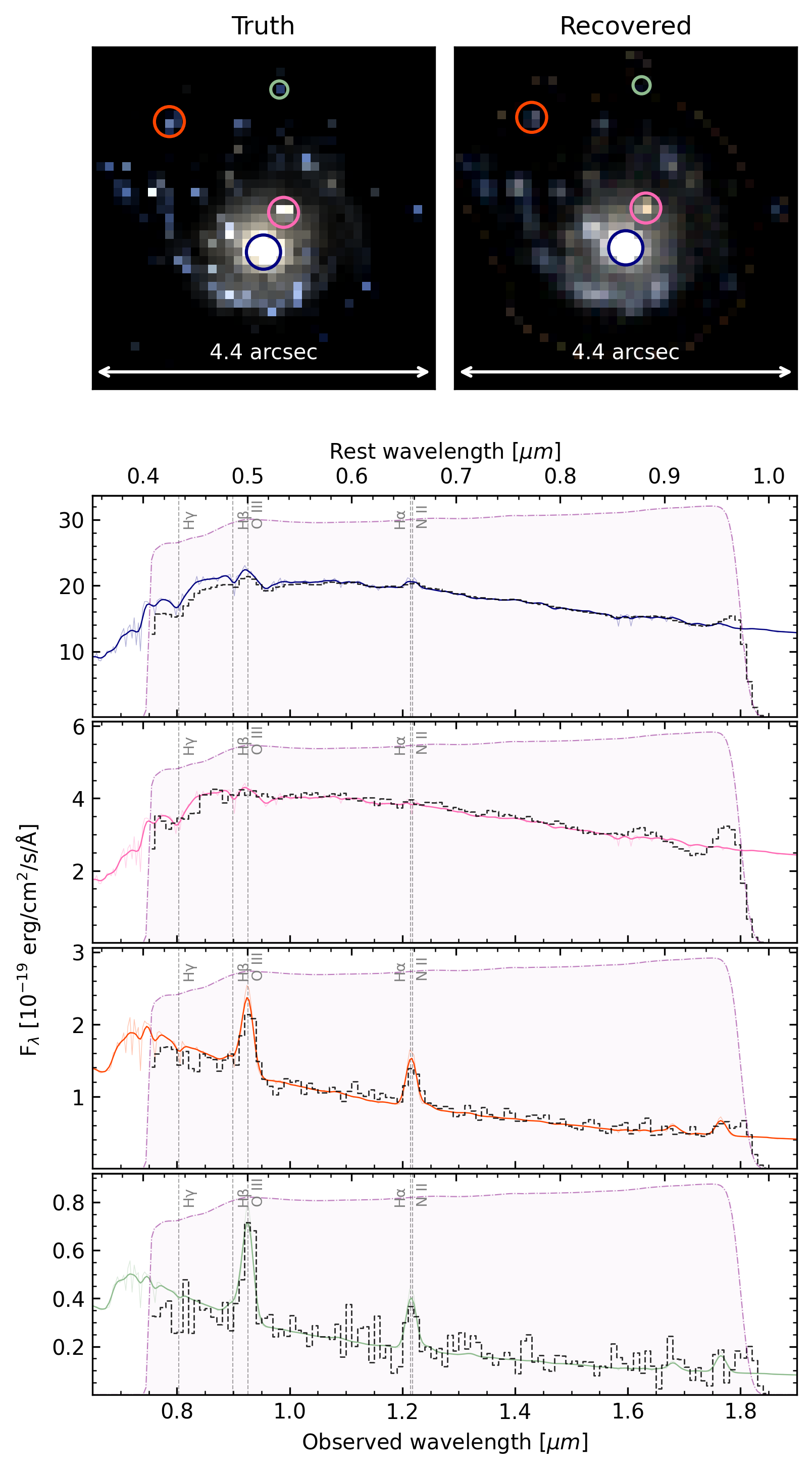}
    \caption{Comparison of input truth and reconstructed flux-cubes at different flux levels. The upper panels display the spatial distribution of the source, showing the RGB image obtained using the true (left) and the recovered (right) flux-cubes. The four colored circles define the extraction apertures. The lower panels show the extracted flux density in units of erg\,cm$^{-2}$\,s$^{-1}$\,${\rm \AA}$$^{-1}$ as a function of observed wavelength (black). Solid lines (color coded as the apertures in the top panels) represent the input spectra convolved with the Roman prism resolution (with the unconvolved spectra in semi-transparent). Key features are identified by vertical dashed lines relative to both observed and rest-frame wavelengths at redshift $z=0.85$. The Roman prism effective area is overlaid in each panel as a purple dash-dotted line to indicate the throughput profile.}
    \label{fig:fc1d}
\end{figure}

\begin{figure}
    \centering
    \includegraphics[width=\columnwidth]{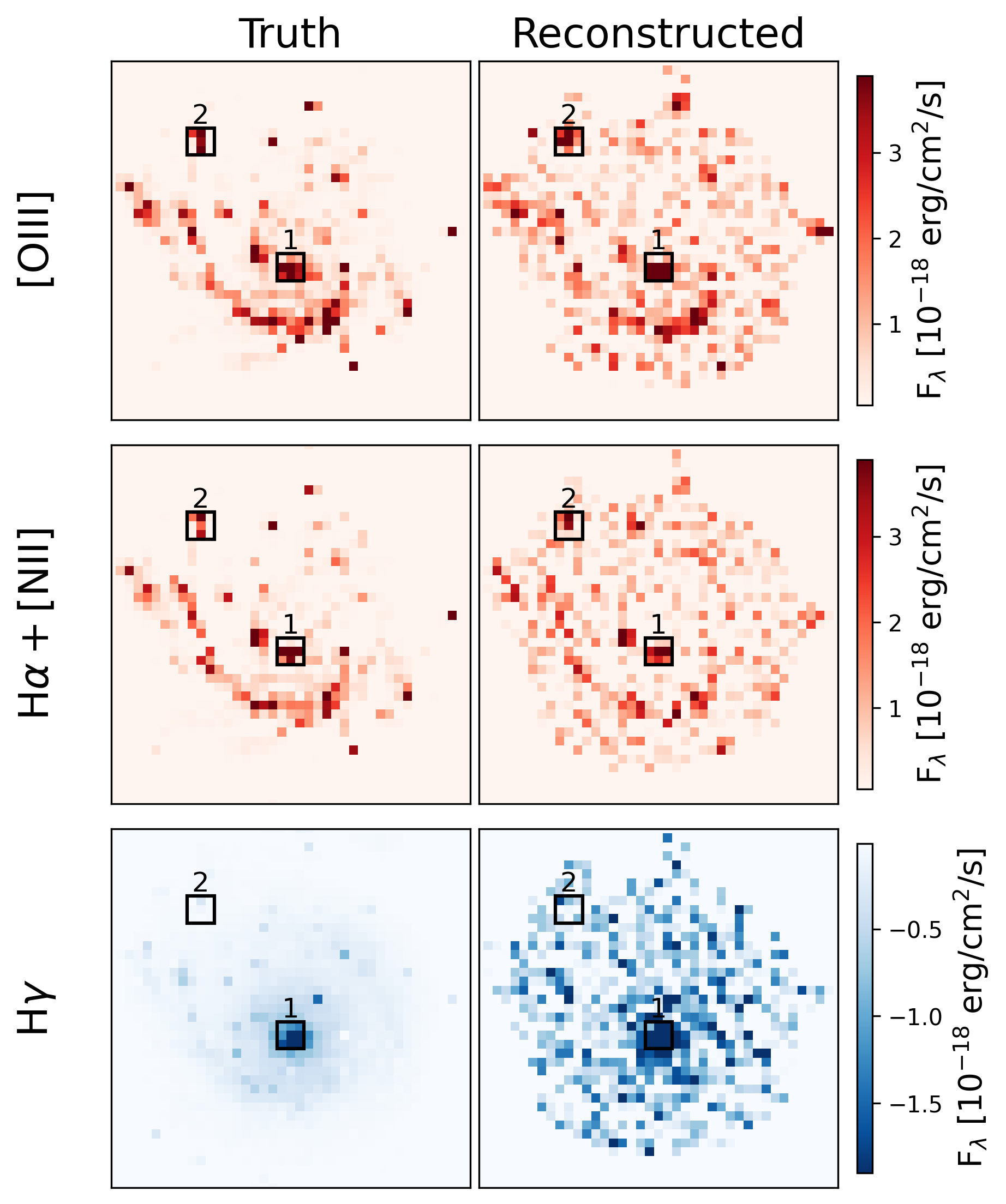}
    \caption{Spatial spectral reconstruction of key lines. The left column shows the resampled input flux-cube, and the right column shows the reconstructed flux-cube. Top and middle rows show the spatial distribution of [\ion{O}{3}] and $H\alpha$ emission, respectively. The bottom row shows the $H\gamma$ absorption profile. A sub-pixel Fourier shift was applied to the truth flux-cube prior to binning to better align it with the spatial grid of the reconstruction and to facilitate visual comparison. The two boxes correspond to the aperture used in the EW measurements of Table\,\ref{tab:line_ew_comparison}.}
    \label{fig:lines}
\end{figure}

Spatially resolved spectroscopy has proven essential for understanding how galaxies assemble their stellar mass, metals, dust, and ionized gas, revealing internal variations in star formation, kinematics, chemical enrichment, and feedback that are inaccessible to integrated measurements alone. Traditionally, such studies have relied on dedicated Integral Field Units (IFUs), which remain observationally expensive and limited to relatively small, targeted samples. Slitless spectroscopy offers a powerful alternative by efficiently sampling large, unbiased galaxy populations, but presents significant challenges for the data analysis.

Although the Roman Space Telescope does not include a dedicated IFU, the HLTDS will deliver an unprecedented slitless-spectroscopy dataset, comprising approximately 72 distinct dispersion angles over 4.5\,deg$^2$ after the completion of the survey \citep{rotac25}. By exploiting this exceptional diversity of roll angles, our reconstruction framework enables an IFU-like recovery of three-dimensional flux distributions, effectively transforming slitless observations into spatially resolved spectral data cubes. 

In this section, we demonstrate the performance of our approach using simulated Roman prism observations of the \texttt{Vela07} galaxy, highlighting its ability to recover spatially resolved emission and absorption features and to enable spatially-resolved studies of star formation and ionized gas across cosmic time. This galaxy is at $z\!=\!0.85$, and was selected for its complex, asymmetric morphology, which provides a stringent test of three-dimensional spatial–spectral recovery. 

Therefore, to emulate the HLTDS, we simulate a dataset that spans 72 unique dispersion angles uniformly sampling 0-360$^\circ$ in steps of 5$^\circ$, with eight dithers per dispersion angle, with a total exposure time of one hour per dispersion angle, consistent with the depth expected for the HLTDS deep spectroscopic tier \citep{rotac25}.

Figures\,\ref{fig:fc1d} and\,\ref{fig:lines} summarize the reconstruction performance. The recovered flux cube reproduces the overall spatial morphology of the galaxy and preserves relative flux variations across distinct regions. Spectra extracted from multiple apertures show good qualitative agreement between the reconstructed and input fluxes across the wavelength range and flux levels, including the recovery of key spectral features such as H$\gamma$, [\ion{O}{3}], and H$\alpha$ at the correct observed and rest-frame wavelengths. The spatial maps of individual lines further demonstrate that both emission ([\ion{O}{3}], H$\alpha$) and absorption (H$\gamma$) features are recovered with the correct spatial distribution and relative intensity. Together, these results show that, given sufficient diversity in dispersion angles and signal-to-noise ratio, the reconstruction algorithm can recover spatially resolved spectral information from slitless data with fidelity comparable to low-resolution IFU observations.

\begin{table}[t]
\centering
\caption{Comparison of Truth and Recovered Equivalent Width (EW) values using a $3\times3$ aperture for Regions 1 (high-flux, low-EW) and 2 (low-flux, high-EW) shown in Fig.\,\ref{fig:lines}.}
\begin{tabularx}{\columnwidth}{lcccc}
\hline
\textbf{Line} & \textbf{Region} & \textbf{Truth} [\AA] & \textbf{Rec.} [\AA] & \textbf{Error (\%)} \\
\hline
\hline
\multirow{2}{*}{[\ion{O}{3}]} &  1 & $22.03$ & $24.15$ & $10\%$ \\
                             &  2 & $158.24$ & $173.93$ & $10\%$ \\
\hline
\multirow{2}{*}{$\mathrm{H}\alpha+$[\ion{N}{2}]} &  1 & $14.30$ & $12.53$ & $-12\%$ \\
                                                &  2 & $150.05$ & $125.15$ & $-17\%$ \\
\hline
\multirow{2}{*}{$\mathrm{H}\gamma$} &  1 & $-12.07$ & $-32.52$ & $>100\%$ \\
                                    &  2 & $-2.68$ & $-20.47$ & $>100\%$ \\
\hline
\end{tabularx}

\label{tab:line_ew_comparison}
\end{table}

We characterized the reconstruction performance by measuring the equivalent width (EW) and its associated percentage error for the three features shown in Fig. \ref{fig:lines}. For both the ground-truth and recovered flux cubes, we extracted spectra from two 3\,$\times$\,3 pixels apertures: Region 1 (the high-flux, low-EW core) and Region 2 (the low-flux, high-EW outskirts).
These regions were selected to test the reconstruction against two competing challenges: signal-to-noise ratio (S/N) and line contrast. In Region 1, the relatively high flux ensures a robust S/N, but the low line contrast makes the feature more susceptible to continuum fluctuations. Conversely, Region 2 offers high line contrast but suffers from a lower S/N due to the faintness of the region.
The EWs were calculated by integrating the flux within the slices corresponding to the spectral lines, while the underlying continuum was estimated from the wavelength slices immediately adjacent to the features.
The EW are reported in Table\,\ref{tab:line_ew_comparison}. For the emission features, the percentage errors relative to the ground truth are approximately 10\% for [\ion{O}{3}] and 10--20\% for H$\alpha$. However, a significant discrepancy arises in the H$\gamma$ absorption feature, where the flux deficit is over-predicted. This error stems primarily from the physical nature of the line: its narrow profile and low contrast against the continuum make it particularly difficult to reconstruct accurately using the low-resolution prism.

In addition to measuring line strengths, we evaluated the spectral fidelity of the reconstruction by determining the galaxy redshift. For this analysis, we focused on the emission features in the high-EW Region 2. Each line was modeled using a Gaussian profile superimposed on a linear continuum to account for the local spectral slope. The fit was performed by varying the line amplitude, centroid, and width, alongside the slope and intercept of the underlying continuum. From these fits, we obtained $z\!=\!0.852\pm0.003$ using the [\ion{O}{3}] line and $z\!=\!0.858\pm0.002$ from H$\alpha$. While the [\ion{O}{3}] estimate matches the ground-truth redshift ($z\!=\!0.852$), the H$\alpha$ result exhibits a minor offset. This discrepancy is likely a consequence of the prism low spectral resolution, which leads to blending with the neighboring [\ion{N}{2}].  Importantly, a more thorough determination of the redshift from these blended lines would account for the known line ratio of [\ion{N}{2}] lines with independent scaling of H$\alpha$/[\ion{N}{2}].  But our present goal is to establish the efficacy of the forward-modeling technique, and leave precise redshift estimation to future work.

As in the previous section, we determined the optimal damping parameter by uniformly spanning a range of values. The best-performing parameter was selected based on a qualitative inspection of spectral fidelity in the four regions of Fig.\,\ref{fig:fc1d}, supported by a simultaneous check of the uncertainties in the calculated EWs. We found that $\log_{10}(\hat{\ell})=-2.8$ provides the best results according to our criteria.

\begin{figure*}
    \centering
    \includegraphics[width=\textwidth]{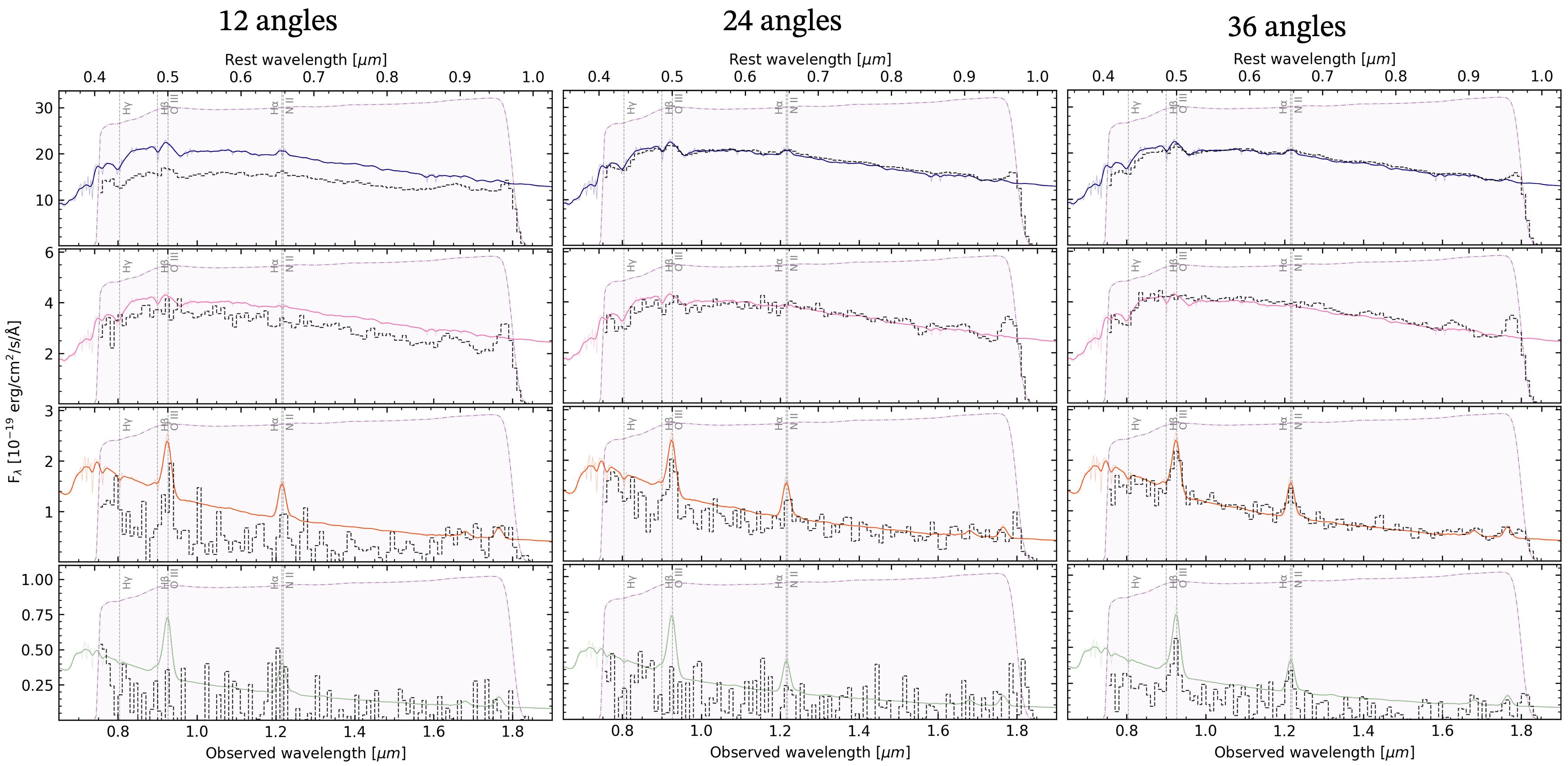}
    \caption{Comparison of spectral reconstructions as a function of angular sampling. Panels follow the layout of Fig. \ref{fig:fc1d} (bottom), illustrating the impact of utilizing 12, 24, and 36 dispersion angles on flux recovery.}
    \label{fig:extr_mult}
\end{figure*}

Finally, to further characterize the reconstruction, we tested the impact of reducing the number of dispersion angles. We performed reconstructions using 12, 24, and 36 angles, each equally spaced across a full 360$^\circ$ rotation. The optimal damping was determined as in the 72-angles case; we found $\log_{10}(\hat{\ell})=-3$ for the 12- and 24-angles cases, and $\log_{10}(\hat{\ell})=-2.8$ for the 36-angles case. 

Figure \ref{fig:extr_mult} displays the extracted spectra for the same four regions introduced in Fig.\,\ref{fig:fc1d}. While the 12-angles reconstruction fails to accurately recover the continuum level, even in high-flux regions, the 24- and 36-angles cases yield results comparable to the full 72-angles set. As shown by the EWs in Table \ref{tab:line_ew_multi}, the 36-angles reconstruction produces errors nearly identical to the 72-angles benchmark, while the 24-angles case struggles on the fainter regions. This suggests that 36 angles provide sufficient sampling; doubling the angle density offers marginal improvements, except in the low-flux regions (bottom panels), where the reconstruction still benefits from the higher total exposure time. However, further testing is necessary to establish the minimum number of dispersion angles necessary as a function of scene complexity and signal-to-noise ratio.

\begin{table}[t]
\centering
\caption{Comparison of Truth and Recovered Equivalent Width (EW) values using a $3\times3$ aperture for Regions 1 and 2 shown in Fig.\,\ref{fig:lines}, but with a different number of dispersion angles used in the reconstruction.}
\begin{tabularx}{\columnwidth}{lcccc}
\hline
\textbf{Line} & \textbf{Region} & \textbf{Truth} [\AA] & \textbf{Rec.} [\AA] & \textbf{Error (\%)} \\
\hline
\hline
\multicolumn{5}{c}{36 angles} \\
\multirow{2}{*}{[\ion{O}{3}]} &  1 & $22.03$ & $22.04$ & $<1\%$ \\
                             &  2 & $158.24$ & $184.73$ & $17\%$ \\
\hline
\multirow{2}{*}{$\mathrm{H}\alpha+$[\ion{N}{2}]} &  1 & $14.30$ & $12.88$ & $-10\%$ \\
                                                &  2 & $150.05$ & $128.62$ & $-14\%$ \\
\hline
\multirow{2}{*}{$\mathrm{H}\gamma$} &  1 & $-12.07$ & $-32.85$ & $>100\%$ \\
                                    &  2 & $-2.68$ & $-3.47$ & $30\%$ \\
\hline\hline
\multicolumn{5}{c}{24 angles} \\
\multirow{2}{*}{[\ion{O}{3}]} &  1 & $22.03$ & $23.18$ & $5\%$ \\
                             &  2 & $158.24$ & $222.53$ & $41\%$ \\
\hline
\multirow{2}{*}{$\mathrm{H}\alpha+$[\ion{N}{2}]} &  1 & $14.30$ & $15.07$ & $5\%$ \\
                                                &  2 & $150.05$ & $65.53$ & $-56\%$ \\
\hline
\multirow{2}{*}{$\mathrm{H}\gamma$} &  1 & $-12.07$ & $-28.47$ & $>100\%$ \\
                                    &  2 & $-2.68$ & $-49.96$ & $>100\%$ \\
\hline
\end{tabularx}

\label{tab:line_ew_multi}
\end{table}

As a final remark, we note that the convergence of the damping parameter across different cases indicates that the scale-invariant normalization successfully identifies a characteristic threshold for suppressing ill-conditioned oscillations of the solution. We find that $\log_{10}(\hat{\ell})\approx-3$ consistently marks the transition where the reconstruction begins to degrade due to noise-driven artifacts, while values $\log_{10}(\hat{\ell})\!\gtrsim\!-2.5$ tend to damp the solution too aggressively, resulting in an over-smoothed reconstruction.

\section{Algorithm implementation and performance}
\label{sec:alg}

\begin{figure} 
    \centering
    \includegraphics[width=\columnwidth]{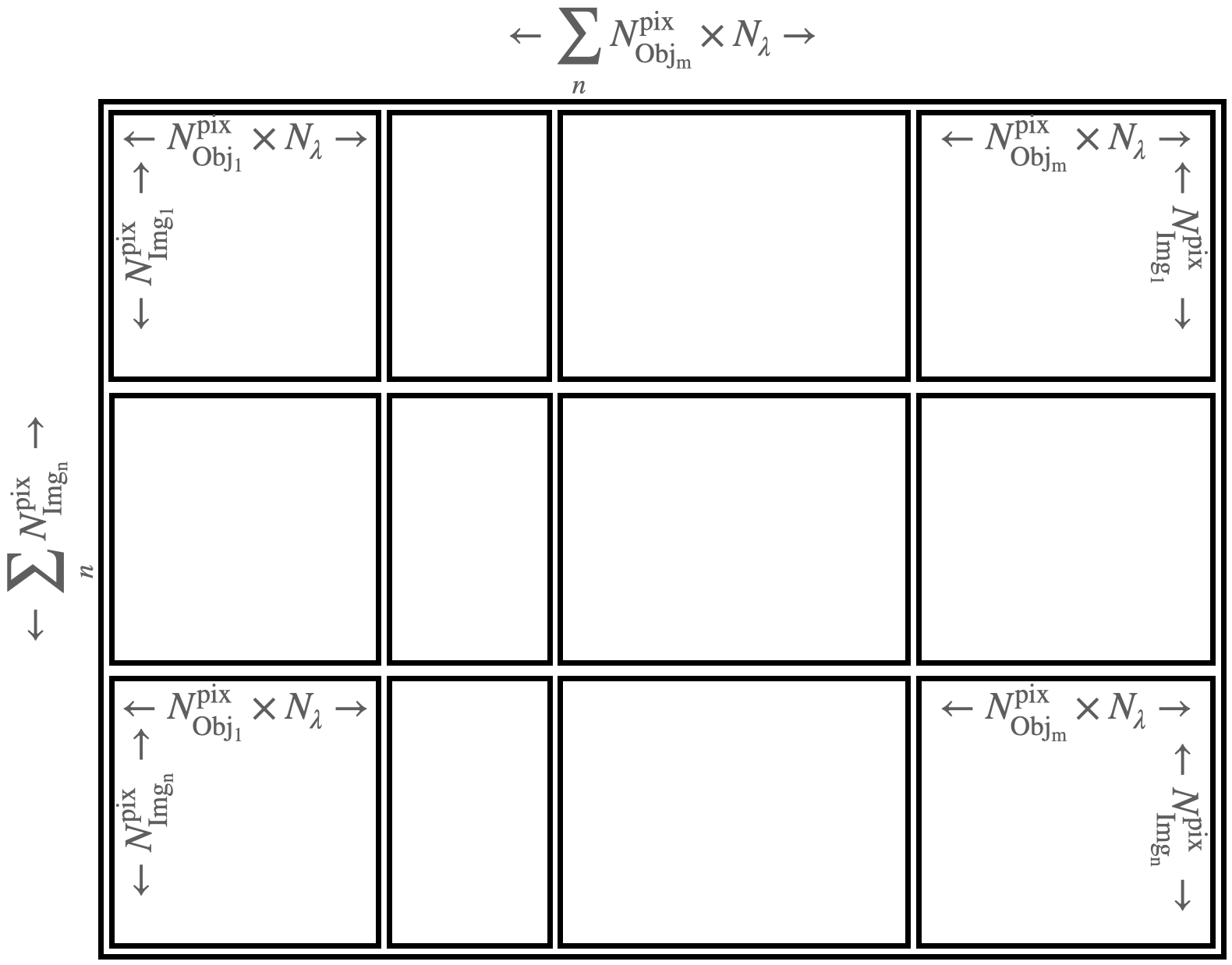}
    \caption{Schematic representation of the linear operator $\mathbf{A}$ used in our algorithm. The operator is organized into $n$ row blocks, each corresponding to an individual input image with $N_{x} \times N_{y}$ pixels. Column-wise, the operator is divided into $m$ blocks representing the $m$ distinct objects being reconstructed. Each individual block is dimensioned by the number of pixels in the $m^{\rm th}$ object ($N^{\rm pixels}_{\rm Obj_{m}}$) and the number of spectral elements ($N_{\lambda}$).}
    \label{fig:linop}
\end{figure}

The implementation of our algorithm is in Python, leveraging high-performance libraries to achieve computational efficiency while maintaining code readability. To ensure near-native execution speeds, all performance-critical routines are implemented using \texttt{Numba} \citep{numba}, a Just-In-Time (JIT) compiler that translates Python functions into optimized machine code. This is complemented by the systematic use of vectorized operations via \texttt{NumPy} \citep{numpy}, which delegates heavy array manipulations to optimized C backends.

The linear operator $\mathbf{A}$ is central to the reconstruction and exhibits a distinct block structure (see Fig.\,\ref{fig:linop}). It has $n$ row blocks, where $n$ is the total number of input images. Each row block corresponds to a single image and contains $N_{x} \times N_{y}$ rows (the number of pixels of the image), and $m$ column blocks, where $m$ is the total number of objects being reconstructed. Each object block in the columns is dimensioned $N_{x} \times N_{y}$ rows by $N^{\rm pixels}_{\rm Obj_{m}} \times N_\lambda$ columns, where $N^{\rm pixels}_{\rm Obj_{m}}$ is the number of pixels of the $m^{th}$ object, and $N_\lambda$ is the number of spectral elements to reconstruct. This structure means that the individual blocks are independent of each other. 

To handle the large-scale nature of the reconstruction, the linear operator $\mathbf{A}$ is stored using the Compressed Sparse Row (CSR) format. This format is specifically optimized for efficient matrix-vector multiplications. To further reduce the memory footprint and eliminate redundant calculations, we implement a pre-processing step to identify and discard empty rows from the matrix, ensuring that only informative data points contribute to the computational load. The computational efficiency is optimized by minimizing the injection of background noise, using a segmentation map derived from deep direct imaging to constrain the reconstruction volume. By limiting the three-dimensional inversion to pixels with significant astrophysical flux, we reduce the dimensionality of the design matrix without sacrificing photometric accuracy, enabling high-fidelity results on a much faster timescales than unconstrained global inversions.

A central component of our framework is a custom \texttt{SciPy} \citep{scipy} \texttt{LinearOperator} abstraction. This structure supports parallelized matrix-vector products ($\mathbf{A}\cdot \mathrm{x}$), distributing the workload across available CPU cores to significantly accelerate convergence. The primary hardware limitation is the available system RAM, which dictates the maximum number of pixels and the total volume of input images that can be processed simultaneously.

In our modeling, we construct a library of synthetic effective PSFs (ePSFs, \citealt{2000PASP..112.1360A}). While we adopt the ePSF formalism to account for a pixel-integrated response, our models are generated from the theoretical optical PSF produced by \texttt{stpsf} \citep{2014SPIE.9143E..3XP} rather than being derived empirically from dithered observations. By pre-integrating the theoretical PSF over the pixel response function of the detector and applying an inter-pixel capacitance kernel, we create a library that captures the sub-pixel structure required for high-fidelity forward modeling. The ePSFs are oversampled by a factor 4.
The use of this pre-calculated ePSF library offers a significant computational advantage: rather than performing expensive two-dimensional convolutions for every wavelength element in every object during the construction of $\mathbf{A}$, the model reduces the operation to a local interpolation problem. Specifically, we implement a biquadratic interpolation scheme. For a given wavelength $\lambda$ and sub-pixel offset ($\Delta x$, $\Delta y$), the algorithm estimates local curvature by evaluating a 4$\times$4 neighborhood of the ePSF grid. This ensures that the PSF profile remains smooth and continuous, capturing sub-pixel variations more accurately than standard bilinear interpolation.
The sparsity of the linear operator $\mathbf{A}$ is directly determined by the spatial footprint of the ePSF. By limiting the biquadratic interpolation to a finite $N_{\rm grid} \times N_{\rm grid}$ area ($N_{\rm grid}$ can be specified as an input), we ensure that the operator remains highly sparse. To achieve execution speeds comparable to compiled C or Fortran, all performance-critical routines are implemented using \texttt{Numba}.

\subsection{Performance}

We report the quantitative performance of the algorithm for the two scenarios previously discussed in Sections~\ref{sec:hostsub} and \ref{sec:resolved} using a workstation with 32 cores and 256\,GB of RAM\footnote{Apple Mac Studio M3 Ultra}. In both scenarios, we used $N_{\rm grid}=9$.

In the four-dispersion-angles reconstruction (\texttt{Vela01}, $\sim$\,500 pixels), the linear operator $\mathbf{A}$ is dimensioned at $256\,932$ rows by $55\,611$ columns, with a 2.25\,GB of memory footprint. The \texttt{LSMR} solver converged in $\sim$\,200 iterations in about 30 seconds, and the total execution time was approximately 2.6 minutes. 

In the higher-complexity case (\texttt{Vela07}, $\sim$\,900 pixels), which used 72 dispersion angles with 8 dithers each, the operator dimensions increase to approximately $(6.0\times 10^{6}) \times (9.8 \times 10^{4})$. In this regime, the matrix density is $0.91\%$, resulting in a memory footprint of 86.8\,GB. The solver reached convergence in 134 iterations over a duration of 8.5 minutes, while the total time for the reconstruction was approximately 74 minutes. This discrepancy indicates that the majority of the computational overhead is concentrated in the initial assembly of the sparse matrix $\mathbf{A}$. However, once the operator is resident in memory, the JIT-optimized \texttt{LinearOperator} abstraction allows for highly efficient matrix-vector products, ensuring that the iterative inversion remains a relatively small fraction of the total processing pipeline. 

To further evaluate the scalability of the framework, we conducted a test by fixing the observation geometry to 30 dispersion angles with 8 dithers each and varying the total number of reconstructed pixels (see Table~\ref{tab:scaling}). The results show that while RAM usage grows linearly with the number of pixels, the execution time does not scale as rapidly. For instance, increasing the pixel count by a factor of $\sim$18 only resulted in a 4.6-fold increase in total time. This indicates that the iterative inversion process remains highly efficient even at larger scales, and the primary time penalty remains the initial construction of the operator $\mathbf{A}$. Consequently, the ability of the framework to handle high-resolution reconstructions is limited more by the available workstation RAM than by the computational cost of the solver itself.

\begin{table}[t]
\centering
\caption{Computational scaling performance for a fixed geometry of 30 angles and 8 dithers.}
\label{tab:scaling}
\begin{tabular}{l c c}
\hline
\hline
Pixel Count & Memory footprint$^{\dagger}$ & Total Execution Time \\
\hline
110   & 4\,GB  & 10 minutes \\
450   & 15\,GB & 18 minutes \\
900   & 40\,GB & 40 minutes \\
1976  & 77\,GB & 46 minutes \\
\hline
\end{tabular}
\begin{flushleft}
{$^{\dagger}${\bf Note:} the current implementation requires twice the reported memory at peak, as it must accommodate the temporary storage needed for full matrix assembly.} 
\end{flushleft}
\end{table}

Given the independent nature of the row blocks in $\mathbf{A}$, the assembly process is naturally suited for high-level parallelization. While the current implementation performs this step sequentially, the total execution time could be further reduced by distributing the construction of individual image blocks across multiple CPUs or compute nodes. Beyond parallelization, the framework could  transition to an ``on-the-fly'' assembly approach. By computing the non-zero elements of the operator dynamically during each iteration rather than pre-calculating and storing the full matrix, the significant RAM overhead identified in Table~\ref{tab:scaling} could be virtually eliminated\footnote{While this strategy solves memory constraints, it is important to acknowledge that on-the-fly assembly typically increases total execution time, as the elements must be recomputed during every iteration rather than accessed from memory.}. Such an enhancement is particularly relevant for highly crowded fields or datasets with large numbers of dither angles, where the number of modeled objects increases. Ultimately, transitioning to on-the-fly assembly would mitigate the current memory and assembly bottlenecks, allowing the framework to scale toward arbitrarily large datasets, dense crowded fields, and higher spatial resolutions that currently exceed workstation memory limits.

\section{Discussion and conclusion}
\label{sec:concl}

In this paper, we presented a novel method for reconstructing three-dimensional flux-cubes of extended sources from slitless spectroscopy data. By leveraging observations acquired at multiple dispersion angles and dithers, our approach recovers the coupled spatial and spectral properties of an object, effectively providing the capabilities of a low-resolution IFU from a slitless spectrograph, albeit with higher noise. Crucially, our framework is entirely non-parametric and data-driven; it does not rely on model libraries or spectral templates, allowing the data to dictate the recovered structure without the constraints of prior assumptions \citep[see, e.g.,][for examples of parametric, template-based methods]{2021ApJ...923..222M,2024MNRAS.532..577E}.

Using simulated \textit{Roman Space Telescope} images, we demonstrated the versatility of this method in two primary applications. First, we showed that by reconstructing a host galaxy model from multiple dispersion angles, we can accurately model the host galaxy and subtract it from the spectral trace of a transient. This approach isolates the transient signal without introducing significant spectral biases, as evidenced by the successful extraction of a Type Ia supernova spectrum. Second, we applied the method to a complex galaxy to recover its three-dimensional spatial-spectral structure. Because the method makes no assumptions about the underlying spectral shape, it is capable of recovering the full range of spectral features, including continuum, emission lines, and absorption, even though narrow absorption features remain challenging at this spectral resolution. Furthermore, we demonstrated that our method does not require a prior redshift to be assumed; rather, it enables the direct measurement of redshifts from the reconstructed flux cubes, for which we found nearly perfect agreement with the ground-truth values.

A key advantage of this framework is its inherent handling of source confusion, which follows directly from the pixel-based discretization of the linear operator. By treating each (sub-)pixel as an independent flux element, the presence of neighboring sources does not introduce systematic modeling errors. Instead, multiple objects are managed as a larger set of parameters to be solved simultaneously within the global linear system. Furthermore, while we demonstrated this technique on Roman simulations, it is fundamentally instrument-agnostic and applicable to any slitless spectrograph, including those on \textit{HST}, \textit{JWST}, and \textit{Euclid}. When applied to wide-area datasets, this method can enable spatially-resolved star formation studies for an unprecedented and unbiased sample of galaxies.

It should be emphasized, however, that the fidelity of the reconstructed flux-cube is coupled to the observational strategy and the mathematical regime of the inversion. For sparse configurations (e.g., four dispersion angles), the framework can find a solution that minimizes the two-dimensional residuals of unobserved projections that are close to the training data, even if the underlying flux-cube is not physically accurate. Conversely, achieving a physically accurate, unbiased three-dimensional reconstruction of a complex, extended source demands a much higher density of angular coverage. Discerning between these two regimes---residual minimization at nearby angles versus a genuinely constrained physical state---is critical when optimizing future survey geometries for high-fidelity three-dimensional reconstruction.

The computational feasibility of quantifying the uncertainties of our estimate depends heavily on the choice of iterative solver. While our baseline framework utilized the \texttt{LSMR} algorithm, which provides efficient convergence but lacks a native mechanism for uncertainty tracking, switching to the \texttt{LSQR} algorithm allows us to extract the variance of each element in the source vector $\mathbf{f}$ with very small additional computational overhead. However, if the system is regularized, the variance computed by \texttt{LSQR} represents the uncertainty of the damped system, capturing how random noise in the observation vector $\mathbf{F}$ propagates into the solution while leaving the systematic bias introduced by the damping factor itself unquantified. In this work, because the true solution was known, we explicitly spanned the damping parameter space to select the optimal threshold. As previously discussed, in general applications where the true solution is unavailable, various parameter-choice techniques can be used to achieve this optimal balance. This ensures the damping is large enough to squash noise-driven variance, yet low enough to preserve the underlying signal and prevent excessive bias, ultimately minimizing the total error of the estimate. We note that a detailed treatment of estimating unbiased errors via a Monte Carlo bootstrap around the damped solution technique has already been thoroughly presented by \cite{ryan18}; given that this foundational groundwork exists, we limit our focus here to the direct algorithmic extraction of the variance components.

Future work will focus on enhancing the computational efficiency of the algorithm by implementing distributed matrix assembly to enable processing on large-scale cluster nodes. We will also perform extractions on increasingly crowded scenes to further characterize the de-blending limits of our method. Finally, we plan to conduct a comprehensive analysis of the reconstruction fidelity as a function of the number of input pixels and dispersion angles to optimize observation requirements for future surveys.

\begin{acknowledgments}
Funding for the Roman Supernova Project Infrastructure Team has been provided by NASA under contract to 80NSSC24M0023.
The material is based upon work supported by NASA under award number 80GSFC24M0006.
L.G. acknowledges financial support from the project PID2023-151307NB-I00 and the Spanish program Unidad de Excelencia María de Maeztu CEX2020-001058-M, financed by MCIN/AEI/10.13039/501100011033, and by the MaX-CSIC Excellence Award MaX4-SOMMA-ICE.
\end{acknowledgments}

\software{\texttt{Astropy} \citep{astropy:2013, astropy:2018, astropy:2022}, \texttt{Matplotlib} \citep{matplotlib}, \texttt{NumPy} \citep{numpy}, \texttt{Numba} \citep{numba}, \texttt{SciPy} \citep{scipy}, \texttt{SNCosmo} \citep{sncosmo}, \texttt{STPSF} (formerly \texttt{WebbPSF}) \citep{per11,per12,per14}.}

\bibliography{bibliography}{}
\bibliographystyle{aasjournal}

\end{document}